\documentclass[12pt,dvips]{article}

\usepackage{amsmath,amssymb,exscale}
\usepackage{array,multicol}
\usepackage{afterpage,float,flafter}
\usepackage{epsfig,rotating,pifont}
\usepackage{cite}
\title{
\vspace{1cm}
\Large\textbf{Tachyons in a slice of AdS}
\vspace*{.5cm}
\author{\large \textbf{
A.~Delgado\footnote{email: adelgado@pha.jhu.edu}
\mbox{  }and M.~Redi\footnote{email: redi@pha.jhu.edu}}\\
\emph{
Department of Physics and Astronomy} \\
\emph{Johns Hopkins University} \\
\emph{3400 North Charles St}. \\
\emph{Baltimore, MD 21218-2686}}}

\date{}
\begin{document}
\maketitle
\thispagestyle{empty}
\vspace*{.5cm}

\begin{abstract}

 Stability of AdS space allows scalar fields to have negative
 mass squared as long as the Breitenlohner-Freedman bound is
 satisfied. In a compactification
 of AdS instead, to avoid instabilities, a tachyonic bulk mass
 must be supplemented by appropriate brane actions.
 In this paper we clarify the meaning of the lower bound in the
 Randall-Sundrum scenario with two branes
 and explain how the instability disappears in the
 infinite space limit. A CFT
 interpretation is also given as radiative symmetry breaking.
\end{abstract}

\newpage
\renewcommand{\thepage}{\arabic{page}}
\setcounter{page}{1}

In the old days of gauged supergravity, it was typically found
that the scalar fields required by supersymmetry have a potential
which is unbounded from below. Anti de Sitter space (AdS) is a
solution of the supergravity equations of motion when the scalars
are at a local maximum of the potential. In flat space this is
clearly unacceptable because the theory would be unstable under
small fluctuations around the vacuum but it was found that this
conclusion does not hold in curved space. In fact, a scalar field
can be stable in AdS space even though the potential is unbounded
from below \cite{freedman}. In particular, a tachyonic mass does
not lead to inconsistencies if the mass satisfies the
Breitenlohner-Freedman (BF) bound which, for the five dimensional
case we will be interested in, is $M^2 \ge -4/L^2$, where $M$ is
the five dimensional mass and $L$ is
the radius of curvature of AdS$_5$. This happens because AdS space
has a boundary where conditions on the fields need to be specified
such that the energy functional is bounded from below even though
the potential is not. In the light of the AdS/CFT correspondence
negative scalars have particular interest because they correspond
to deformations of the CFT by relevant operators (see
\cite{maldacena}). In the CFT language the lower bound on the mass
is understood as a unitary bound on the dimensions of the gauge
invariant operators.

In this brief paper we will study the r\^ole of negative mass
squared scalars in the Randall-Sundrum compactification of AdS and
propose a new holographic interpretation. We became aware that
this subject has been previously investigated in \cite{japanese}
with which our work has some overlapping. In our case we consider
in detail the two brane scenario and show that a tachyonic bulk
scalar naturally leads to instabilities (at least as long as
gravity is non dynamical) for any distance between the two branes.
By studying the limit where one of the branes is sent to the
boundary of AdS we explain how the BF bound arises. Also, our
AdS/CFT interpretation is completely different.

In order to be specific, we consider the compactification of five
dimensional AdS space which has drawn a lot of attention for its
relevance for phenomenological applications, but we believe that
our results hold in any number of dimension with minor
modification of the formulas.
 As shown in \cite{RS1}, for the
particular case of five dimensions, it is possible to compactify
AdS space onto an orbifold $S^1/\mathbb{Z}_2$. This geometry is
obtained by adding two 3-branes parallel to the boundary of AdS to
reproduce the singularities of the metric at the fixed points of
the orbifold. Once the tensions of the branes are tuned with the
cosmological constant of the bulk, this configuration is an exact
solution of Einstein's equation for any distance between the two
branes. This space has a four dimensional ground state which is
ordinary flat space. Our strategy will therefore be to perform a
Kaluza-Klein  (KK) decomposition of the fields and use ordinary
flat space arguments to discuss stability.

We consider the AdS metric in Poincar\'e coordinates:
\begin{equation}
ds^2=\frac {L^2} {z^2} (\eta_{\mu \nu}dx^\mu dx^\nu-dz^2)
\label{metric}
\end{equation}
where $L$ is the radius of curvature of AdS. The space has a
boundary at $z=0$ while $z=\infty$ is the horizon. The physical
space of the orbifold is an interval with metric given by
(\ref{metric}) bounded by two branes located at $z_1$ (UV brane)
and at $z_2$ (IR brane).

 The action of a massive
scalar propagating in this geometry, including brane localized
mass terms is:
\begin{equation}
S=S_{bulk}+S_{UV}+S_{IR}
\label{action}
\end{equation}
where:
\begin{eqnarray}
S_{bulk}&=&\int d^4x \int dz \sqrt{G}(\frac 1 2  G^{M N}
\partial_M \phi
\partial_N \phi-\frac 1 2 M^2 \phi^2)\nonumber\\
S_{UV}&=&-\int d^4x \sqrt{-g_{UV}} \frac {\lambda_1} {L} \phi^2\nonumber
\\
S_{IR}&=&-\int d^4x \sqrt{-g_{IR}}\frac {\lambda_2} {L} \phi^2
\end{eqnarray}
$g_{UV}$ and $g_{IR}$ are the induced metrics on the branes and
$\lambda_{1,2}$ are dimensionless couplings. In general, brane
localized kinetic terms can also be included and they naturally
arise from the dynamic of the system in the interacting theory,
but they will not be relevant for the present discussion. In the
infinite space limit, boundary conditions on the fields have to be
imposed at $z=0$. In the finite slice, the boundary conditions at
$z_1$ and $z_2$ are implied by the brane actions.

The equation of motion following from (\ref{action}) is:
\begin{equation}
\frac {L^3} {z^3}(-\partial^2_z + \frac 3 z \partial_z + \eta^{\mu
\nu}\partial_\mu
\partial_\nu+ \frac {M^2L^2} {z^2}+ \frac {2\lambda_1} {z} \delta(z-z_1)+
\frac {2\lambda_2}
{z}
\delta(z-z_2))\phi(q,z)=0
\end{equation}
The solutions of the bulk equation are linear combinations of
Bessel functions of order $\alpha=\sqrt{4+M^2 L^2}$:
\begin{equation}
\phi(q,z)=z^2(A~J_\alpha(q~z)+B~Y_\alpha(q~z))
\label{wavesolutions}
\end{equation}
where $q$ is the four dimensional momentum. The delta Dirac impose
the following boundary conditions on the fields:
\begin{eqnarray}
&z\partial_z \phi(q,z_1)&=\lambda_1 \phi(q,z_1) \nonumber\\
&z\partial_z \phi(q,z_2)&=-\lambda_2 \phi(q,z_2)
\label{boundaryconditions}
\end{eqnarray}

We have here assumed that $\phi$ is even under the $\mathbb{Z}_2$
projection. For an odd scalar, we should impose that the
wave function vanishes at the orbifold fixed points. This can be
considered as a special case of Dirichlet boundary conditions
which can always be imposed.
 Substituting (\ref{wavesolutions}) into
 (\ref{boundaryconditions})
 and using the properties of the Bessel functions, we obtain an
 equation for the $4D$ KK masses $m$, namely:
\begin{equation}
\frac {(2-\alpha-\lambda_1)J_\alpha(mz_1)+ mz_1 J_{\alpha -
1}(mz_1)} {(2-\alpha-\lambda_1)Y_\alpha(mz_1)+ mz_1 Y_{\alpha -
1}(mz_1)}=\frac {(2-\alpha+\lambda_2)J_\alpha(mz_2)+ mz_2
J_{\alpha - 1}(mz_2)} {(2-\alpha+\lambda_2)Y_\alpha(mz_2)+ mz_2
Y_{\alpha - 1}(mz_2)} \label{kkmasses}
\end{equation}

Due to the reality of the action, solutions to this equation can
either be purely real or purely imaginary. For $\lambda_{1,2}=0$,
corresponding to ordinary Neumann boundary conditions, we find
that the spectrum of the KK masses is non tachyonic if and only if
$M^2 \ge 0$. In particular, for $M^2\ge -4/L^2$ (corresponding to
the BF bound in $5D$), there is no massless mode in the spectrum
unless the bulk mass is zero. Therefore, a scalar propagating in
the RS background is unstable whenever $M^2<0$ as in flat space.
Of course, it is always possible to tune the brane masses to
compensate for the negative bulk mass but this can be done
independently of the BF bound, so it is not different from a
tachyon in flat space. In \cite{pomarol} it was considered the
case of supersymmetric theories in a slice of AdS$_5$. SUSY in AdS
background requires some of the scalars to have negative mass
satisfying the bound. By ordinary $4D$ flat space supersymmetry
all the KK masses have to be positive so it was found that
invariance of the action under supersymmetry transformation
requires $\lambda_{1,2} \ne 0$. In general the brane terms will
depend on the position of the two branes. This is clearly
unsatisfactory, especially in a theory where gravity is dynamical
since in this case the distance between the two branes is free to
fluctuate.

It is natural to ask what is meaning of the lower bound in a slice
of AdS. For zero brane actions we have checked numerically that a
negative mass scalar with $M^2\geq-4/L^2$ always generates a
single $4D$ tachyon plus positive excitations, independently of
the location of the two branes. If we now send the UV brane toward
the boundary, the mass of the tachyon grows to infinity so
effectively this mode decouples from the theory. It will be
convenient for us to keep the UV brane fixed and move the IR brane
to the horizon of the space at $z=\infty$. Appropriately Weyl
rescaling the four dimensional metric will be physically
equivalent to moving the UV brane while keeping the IR brane
fixed. In the limit $z_2\to \infty$, the positive modes become
continuous while the tachyon has a finite mass squared of the
order of $-1/z_1^2$. Technically this happens because, for
imaginary values of $m$ and $|mz_2| >> 1$, equation
(\ref{kkmasses}) becomes:
\begin{equation}
(2-\alpha)H^1_\alpha(m z_1)+ m z_1 H^1_{\alpha - 1}(m z_1)=0
\label{tachyon}
\end{equation}
where $H^1_\alpha(m z) =J_\alpha(m z)+i Y_\alpha(m z)$. From
(\ref{tachyon}) we see that the properties of the tachyon are
associated with the UV brane. This type of equation has been
widely studied in the contest of general properties of Bessel
functions~\cite{bessel} and it can be proved analytically that
equation (\ref{tachyon}) has indeed zero imaginary solutions when
$M^2\ge 0$ and two conjugate when $-4/L^2 \le M^2 <0$ which fully
agrees with our previous numerical checks.

 We
can Weyl rescale the four dimensional metric according to:
\begin{equation}
g_{\mu \nu} \to (z_2/L)^2g_{\mu \nu}
\label{Weyl}
\end{equation}
to obtain the configuration where we keep the IR brane fixed
(where the $5D$ metric has canonical normalization) and the UV
brane is located at $z_1=L^2/z_2$ (assuming that the UV brane
was originally at $z_1=L$). It follows from (\ref{Weyl}) that the
mass of the tachyon is proportional to $1/z_1$ for small $z_1$.

In the limit where the UV brane has been moved all the way to the
boundary of AdS, the equation (\ref{kkmasses})  for the positive
modes also simplifies:
\begin{equation}
(2-\alpha)J_\alpha(m z_2)+m z_2 J_{\alpha-1}(m z_2)=0
\end{equation}
The mass gap between the positive KK modes remains finite and the
mass scale is set by $1/z_2$. Basically, the different behavior of
the tachyon stems from the fact that its wave function is
localized on the UV brane while all the other excitations are
peaked on the IR brane. As a consequence, the two types of modes
decouple from each other when $z_1 \to 0$ and the tachyon is
projected out of the physical spectrum.

We are  now ready to show the inconsistency that arise when $M^2<
-4/L^2$. Moving the IR brane to infinity one finds that additional
tachyons are generated. Actually, equation (\ref{tachyon}) has now
an infinite number of imaginary solutions approaching zero. This
clearly leads to an instability. After Weyl rescaling according to
(\ref{Weyl}), each mode has a mass which grows as $1/z_1$ but, due
to the new tachyons that are generated, there are always states
with $m^2=- O(1/L^2)$. Even if we start with a localized mass term
on the UV brane such that the spectrum is tachyon free, by moving
the UV brane toward the boundary we will eventually generate
tachyons since the required mass grows as $1/z_1$.

One of the features of these scenarios is that there is an AdS/CFT
duality that relates the $5D$ warped theory to a $4D$ CFT (see
\cite{arkani}). The standard correspondence is that a theory on a
slice of AdS$_5$ is dual to a $4D$ CFT with a cut-off given by the
position of the UV brane and in which conformal invariance is
spontaneously broken at a scale set by the IR brane. This last
fact explains the existence of a mass gap between the KK modes in
the case where the IR brane is kept at a finite distance and the
UV brane is removed. Fields living in the bulk of the extra
dimension correspond to fundamental $4D$ fields living outside the
CFT and coupled to gauge invariant operators of the CFT. In the
case of scalars  the conformal dimension of the operator is
related to the bulk mass of the scalar by:
\begin{equation}
\Delta_\pm=2\pm \sqrt{4+M^2 L^2} \label{dimension}
\end{equation}
where the minus sign (necessary to saturate the unitarity bound
$\Delta \ge 1$ of $4D$ CFT) is allowed only in the region
$-4/L^2<M^2<-3/L^2$. \footnote{We would like to thank Prof. Igor
Klebanov for clarifying this point to us.} With our boundary
conditions the positive modes grow as $z^{\Delta_+}$ for small
$z$, in the limit where the UV brane is removed, which corresponds
to the choice of the plus sign in the formula (\ref{dimension})
\cite{klebanov}.  The correspondence also asserts that classical
computations on the AdS side correspond to large-$N$ contributions
in loops of CFT.

The interpretation of the appearance of a tachyon mode when this
kind of scalars is compactified is as follows. In the CFT dual a
mass term localized on the UV brane corresponds to adding a mass
term for this fundamental scalar as well as a deformation of the
CFT, while a mass term on the IR brane is interpreted as a
threshold effect when the CFT is broken. In the case when no brane
term is added there is a massless scalar at the UV scale which is
coupled to a CFT through a relevant operator (i.e. $\Delta<4$),
and the large-$N$ loop contributions drive that scalar to have a
negative mass in the IR. If some brane term is added on the AdS
side, to cancel the negative mass as explained before, then the
holographic interpretation is that there is a massive scalar
in the UV and the CFT does not have enough room to make it
tachyonic. Since the value of the brane term, or UV mass, depends
on the size of the bulk there is a way to go from a healthy scalar
to a tachyon just by changing the position of the UV brane, that
is by changing the scale of the cut-off.
 This means that depending on the scale where the CFT is
spontaneously broken the scalar will have positive or negative
$m^2$. In other words, when this scalar is charged under some
gauge group, this resembles the case of radiative symmetry
breaking, i.e., a perfectly massive scalar in the UV whose mass
is driven negative in the IR by CFT loops. In order for this scalar
to have phenomenological interest for electroweak symmetry
breaking the UV scale should not be much separated from the IR
scale, because as has been shown, the negative mass of the tachyon
is of the order of the UV cut-off, so unfortunately this mechanism
can not be used in the usual RS1 set-up with the Higgs field in
the bulk. One can also understand why in the case when the UV
brane is sent to the boundary the tachyonic mode decouples. The
dual interpretation of that is simply that the UV brane sets the
cut-off of the CFT, so when there is no UV brane there is no
cut-off and the external fields added became non-dynamical so this
potential tachyon decouples from the rest of the fields.

To conclude we can summarize the results of this paper. We have
studied the behavior of tachyonic scalars in the Randall-Sundrum
scenario with two branes (UV/IR), the reason for that being that
full AdS allows that kind of scalars as long as the mass satisfies
the Breitenlohner-Freedman bound. It has been shown that, in
general, those scalars will generate instabilities in a slice of
AdS, unless suitable brane actions are included. It has also been
pointed out that, as long as the bound is satisfied, when the UV
brane is sent to the boundary then the instabilities disappear.
This is very well understood from the KK decomposition because
there is a single tachyonic mode whose wave function is localized
near the UV brane while the positive modes are peaked on the IR
brane. Finally a CFT interpretation has been proposed relating the
generation of this tachyon mode with large-$N$ loops of the CFT.
Any possible phenomenological application or the extension of this
analysis to other geometries is left for future investigation.

\section*{Acknowledgments}
We thank Jon Bagger and Raman Sundrum for useful discussions and
for reading the manuscript. A.~D.~ is supported by NSF Grants
P420D3620414350 and P420D3620434350.

\end{document}